\documentclass[pre,showpacs,amsmath,amssymb,superscriptaddress,twocolumn,longbibliography]{revtex4-1}
\usepackage{graphicx}
\usepackage{dcolumn}
\usepackage{bm}
\usepackage[latin1]{inputenc}
\usepackage{epstopdf}
\usepackage{amsmath}
\usepackage{amssymb}
\usepackage{color}
\usepackage{latexsym}
\usepackage{mathrsfs}

\begin{document}

\title{Slow dynamics and subdiffusion in a non-Hamiltonian system with
long-range forces}

\author{Romain Bachelard}
\affiliation{Departamento de Fisica, Universidade Federal de S\~{a}o
Carlos, Rod. Washington Luis, km 235, S/n - Jardim Guanabara, S\~{a}o
Carlos - SP, 13565-905, Brazil}
\email{bachelard.romain@gmail.com}
\author{Nicola Piovella}
\affiliation{Physics Department, Universit\`{a} degli Studi di Milano,
Via Celoria, 16 - 20133 Milano, Italy}
\email{Nicola.Piovella@mi.infn.it}
\author{Shamik Gupta}
\affiliation{Department of Physics, Ramakrishna Mission Vivekananda
University, Belur Math, Howrah 711202, India}
\email{shamik.gupta@rkmvu.ac.in}

\begin{abstract}
Inspired by one--dimensional
light--particle systems, the dynamics of a non-Hamiltonian system with
long--range forces is investigated. While the molecular dynamics does not reach an
equilibrium state, it may be approximated in the thermodynamic limit by
a Vlasov equation that does possess stable stationary solutions. This
implies that on a macroscopic scale, the molecular dynamics evolves on a
slow timescale that diverges with the system size. At the
single-particle level, the evolution is driven by incoherent interaction
between the particles, which may be effectively modeled by a noise,
leading to a Brownian-like dynamics of the momentum. Because this self-generated diffusion
process depends on the particle distribution, the associated
Fokker-Planck equation is nonlinear, and a subdiffusive behavior of the
momentum fluctuation emerges, in agreement with numerics.
\end{abstract}

\maketitle

Long--range interactions are present at all scales, from atomic physics
to astrophysics, from hydrodynamics to plasma and free-electron laser physics~\cite{Campa2014}.
The lack of additivity of long-range systems challenges several
important results of equilibrium statistical physics found in classical
textbooks and developed for short--range interactions. The most
fundamental consequences are the possibility of a non-concave
entropy~\cite{Touchette2008} and inequivalent microcanonical and canonical ensembles~\cite{Touchette2015}.

It is probably when out of equilibrium that long--range
systems revealed most surprises, with the rather intriguing and
interesting property that the time to reach equilibrium may diverge with the
system size~\cite{Yamaguchi2003,RochaFilho2014}. Coined
quasi--stationarity, this peculiar behavior was shown to derive from the
existence of the so-called Vlasov equation describing the phase-space
dynamics in the thermodynamic limit, which admits a continuum of stable
stationary solutions~\cite{Yamaguchi2004}. An important consequence is
that large systems may essentially remain trapped in out-of-equilibrium
states for times accessible to experiments. These results, obtained for
energy-conserving Hamiltonian dynamics, were nevertheless contrasted
by studies of dynamics that violates energy conservation, in which
stochastic terms were shown to put a bound on the lifetimes of the
out-of-equilibrium states~\cite{Gupta2010a,Gupta2010b,Chavanis2011}.
Nevertheless, until now, the Hamiltonian dynamics has been the
main framework to study the phenomena of quasi--stationarity, as an heritage of statistical physics.

In this Rapid Communication, we show that {\it non-Hamiltonian} systems with
long--range forces may also exhibit quasi-stationary features, despite not ever
reaching an equilibrium. The model under consideration, which may be
achieved either in cold atom or free-electron laser setups, has an
ever-growing kinetic energy. We show that the existence of a general
condition for the stability of stationary solutions of the associated Vlasov
equation allows for the presence of quasi-stationary states. For
non-magnetized states, each particle is driven by a fluctuating
magnetization that can effectively be modeled as a stochastic noise,
which in turn allows to derive a nonlinear Fokker-Planck equation for
the momentum distribution.
Assuming that the system reaches a Gaussian distribution in momentum, a
subdiffusive behavior of momentum fluctuations is predicted, in
agreement with our numerical findings. Our work reveals a surprising
dynamical possibility allowed by non-Hamiltonian long-range forces. Thermodynamically, the system does not have a
long-time equilibrium stationary state to relax to. Nevertheless,
dynamically, the system remains trapped in states for times that diverge
with the system size, so that such states become in the limit of large
system size the effective stationary states of the system. This work is to the best
of our knowledge the first demonstration of quasi--stationarity in
non-Hamiltonian long-range systems. We also offer possible experimental
platforms to observe our predicted findings. 

The physical model we consider here is the one-dimensional dynamics of
particles interacting with light, as may be achieved in free-electron
laser~\cite{Bonifacio1985} and cold atom~\cite{Bonifacio1997} set-ups.  
In these systems, the particles typically behave as pendula coupled by
the common radiation field. For example, a cloud of cold atoms in a ring optical cavity backscatters the photons from an incident pump beam into a counter-propagating cavity mode, according to the following equations:
\begin{subequations}
\begin{eqnarray}
 \dot\theta_j &=& p_j,~~\dot p_j = -g(Ae^{i\theta_j}+\mathrm{c.c.}),\label{pj}\\
 \dot A &=& \frac{g}{N}\sum_{j=1}^N e^{-i\theta_j}-(\kappa-i\Delta)A,\label{A}
 \end{eqnarray}
\end{subequations}
where $\theta_j$, $p_j$ and $A\propto 1/\sqrt{N}$ are respectively the normalized
positions and the momenta of the $N$ particles and the cavity field
amplitude, while c.c. stands for complex conjugate. Here,
$g\propto\sqrt{N}$ describes the coupling  between the atoms and the
field~\cite{Bonifacio1985,Bonifacio1997}, $\kappa$ models the cavity
losses and $\Delta$ is the frequency mismatch between the cavity and the
atomic transition. The $1/N$ term in Eq.~\eqref{A} allows considering
the thermodynamic
limit ($N\to\infty$) of the problem without encountering divergences, in
accordance with the Kac prescription~\cite{Kac1963}.

For a bad--quality mirror ($\kappa> g^{2/3}$), the scattered field
quickly leaves the interaction region, while the atoms continuously lose
momentum by emitting photons into the cavity mode.
The system then enters into a superradiant regime in which the atoms
scatter a transient radiation pulse with intensity proportional to
$N^2$. The same regime can be achieved in a free-electron laser
operating with short electron bunches~\cite{SR:FEL}. The adiabatic
elimination of the field amplitude reads $
A\approx g/(\kappa-i\Delta)\sum_{j=1}^N e^{-i\theta_j}/N$,
which in turn leads to the following equations:
\begin{eqnarray}
\dot{\theta}_j &=&p_j,~~\dot{p}_j=-\frac{2g^2\kappa}{\kappa^2+\Delta^2}
\frac{1}{N}\sum_{m=1}^N \cos(\theta_j-\theta_m)\nonumber
\\ &&\hspace{1.7cm}+\frac{2g^2\Delta}{\kappa^2+\Delta^2} \frac{1}{N}\sum_{m=1}^N
\sin(\theta_j-\theta_m).\label{eq:pjdot0}
\end{eqnarray}
For light far-detuned in the blue ($\Delta\gg\kappa$), the cosine term
in the second equation may be dropped, and one recovers a Hamiltonian
dynamics that has been studied extensively under the name of the
Hamiltonian Mean-Field Model~\cite{Antoni1995}.
On the contrary, at resonance ($\Delta=0$), the dynamics is strongly
dissipative, a case on which we focus from now on. Also, since it
corresponds to a rescaling of time and momentum, we set from now on
$2g^2\kappa/(\kappa^2+\Delta^2)=1$ without loss of generality.

The macroscopic ordering of the particles is captured by the magnetization $M \equiv
(1/N)\sum_{j=1}^N e^{-i\theta_j}$ that may be used to rewrite the
dynamical equations as
\begin{equation}
\dot{\theta}_j=p_j,~~
\dot{p}_j=-\frac{1}{2}\Big(Me^{i\theta_j}+M^\ast e^{-i\theta_j}\Big).
\label{eq:pjdot}
\end{equation}
An important feature is that the force
$F_{jm}=-(1/N)\cos(\theta_j-\theta_m)$ on particle $m$ due to particle
$j$ does not have the symmetry of a
force derivable from a two--body interaction potential that is a function
solely of the separation between particles. In the latter
case, one has $\mathbf{F}_{jm}=-\mathbf{F}_{mj}$, which is the situation typical of Hamiltonian
systems encountered in statistical mechanics, and which ensures that the value
of the average
momentum $P\equiv (1/N)\sum_{j=1}^N p_j$ is conserved in time. The
dynamics~\eqref{eq:pjdot} is not derivable from an underlying
Hamiltonian, so that one may not associate an energy function with the
system. The average momentum for our model is not conserved but instead decreases in time according to
\begin{equation}
\dot{P}=-|M(t)|^2.
\end{equation}
Even in a non--magnetized phase, while $M(t)$ averages to zero over time,
the fluctuations of $|M|$ will contribute to the decrease of the total momentum. Consequently, the system does not possess
a proper equilibrium, with a momentum distribution that
is stationary in time. 

\begin{figure}[!ht]
\centering
\includegraphics[width=8cm]{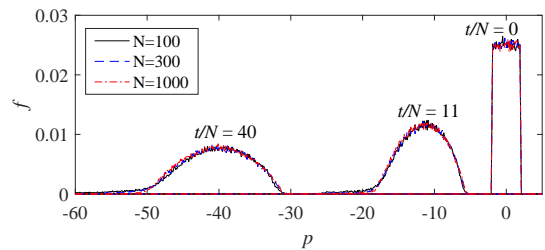}
\caption{(Color online) The single-particle $p$-distribution as a
function of rescaled time $t/N$ for
the dynamics~\eqref{eq:pjdot} for three different system size $N$ and for
an initial state that is WB with $\Delta p=0.5$. The data are
obtained from numerical integration of the dynamics~\eqref{eq:pjdot} for
different system sizes.}
\label{fig:p-profiles}
\end{figure}

The decrease of $P$ with time is confirmed by
numerical simulations of the dynamics~\eqref{eq:pjdot}, as
may be concluded from Fig.~\ref{fig:p-profiles} by observing the shift of the centre of the momentum
distribution and the collapse of the curves for different system size
$N$ on scaling time by $N$. The latter observation implies a rather strong
dependence of the dynamics on the system
size $N$, suggesting a slowing down of the evolution with increase of
$N$. Similar slowdown of macroscopic evolution in systems with
long-range interaction has already been reported for Hamiltonian
dynamics~\cite{Campa2014,Gupta2017}, and may be explained as resulting
from the occurrence of a continuum of stable stationary solutions of the
Vlasov equation describing the macroscopic evolution of the system in
the thermodynamic limit~\cite{Campa2014,Bouchet2005}.

Although our model~\eqref{eq:pjdot} is intrinsically non--Hamiltonian, it
is instructive, especially in the light of our observation of slow
relaxation mentioned above, to derive a Vlasov equation to describe
its dynamics in the limit of large $N$. To this end, let us introduce
the single--particle
density $f_d(\theta,p,t)\equiv
(1/N)\sum_{j=1}^N\delta(\theta-\theta_j(t))\delta(p-p_j(t))$ as the
density of particles with angle $\theta$ and momentum $p$ at time $t$.
Taking the time derivative of $f_d$ and using the equations of
motion~\eqref{eq:pjdot}, it may be shown that in the limit of large $N$, when the discrete function $f_d(\theta,p,t)$ approaches
a continuous one, namely, the single-particle distribution function
$f(\theta,p,t)$, the time evolution of the latter is given by a
Vlasov equation of the following form (for the general procedure,
see Ref.~\cite{Campa2014}):
\begin{equation}
\frac{\partial f}{\partial t}+p\frac{\partial f}{\partial
\theta}+F[f](\theta,t)\frac{\partial f}{\partial p}=0.
\label{eq:Vlasov}
\end{equation}
Here, $F[f](\theta,t) \equiv -\iint {\rm d}\theta' {\rm
d}p'~f(\theta',p',t)\cos(\theta-\theta')$, a functional of $f$, is the
net force experienced by a particle with angle $\theta$ at time $t$; 
$f(\theta,p,t)$ obeys the
normalization $\iint {\rm d}\theta {\rm
d}p~f(\theta,p,t)=1\ \forall~t$; the magnetization is given by $M[f](t)=\iint {\rm d}\theta {\rm
d}p~f(\theta,p,t)e^{-i\theta}$.

The stationary states of Eq.~\eqref{eq:Vlasov} satisfy $\partial f_{\rm
s}/\partial t=0$. Let us focus on non-magnetized stationary states, which
correspond to $F[f_{\rm s}]\equiv 0$, so that any state $f_0(p)$ that is
homogeneous in $\theta$ is a stationary solution of
Eq.~\eqref{eq:Vlasov}. Its linear stability is determined by considering
the expansion $f(\theta,p,t)=f_0(p)+\delta f(\theta,p,t)$, with $\delta f$ an eigenvector of the linearized dynamics whose norm satisfies $||\delta
f(\theta,p,t)||\ll 1$, so that inserted in Eq.~\eqref{eq:Vlasov}, one
obtains to leading order the equation
\begin{equation}
\frac{\partial \delta f}{\partial t}+p\frac{\partial \delta f}{\partial
\theta}+F[\delta f](\theta,t)f_0^\prime(p)=0,
\label{eq:Vlasov-linearized}
\end{equation}
where the prime denotes the derivative. 
Using the fact that $\delta f$ is $2\pi$-periodic in $\theta$, we expand
the perturbation $\delta f$ as $
\delta f(\theta,p,t)=\sum_{k=-\infty}^\infty \widetilde{\delta
f}_k(p)e^{ik\theta+\lambda t}$;
$\delta f$ being real implies that $\widetilde{\delta f}_{-k}=\widetilde{\delta
f}_k^\ast$. We then have $F[\delta
f](\theta,t)=-\pi \int {\rm d}p'~\widetilde{\delta
f}_k(p')e^{ik\theta+\lambda t}(\delta_{k,1}+\delta_{k,-1})$. On substituting this expression in
Eq.~\eqref{eq:Vlasov-linearized}, we find that the Fourier coefficients
$\widetilde{\delta f}_{\pm 1}$ satisfy the equation $
\widetilde{\delta f}_{\pm 1}(p)=\pi f_0^\prime(p)/(\lambda \pm
ip)\int {\rm d}p'~\widetilde{\delta
f}_{\pm 1}(p')$.
On integrating both sides with respect to $p$ and noting that $\int {\rm
dp}~\widetilde{\delta f}_{\pm 1}(p) \ne 0$, one gets the dispersion
relation determining the stability parameter $\lambda$:
\begin{equation}
1=\pi \int{\rm d}p~\frac{f_0^\prime(p)}{\lambda \pm i p}.
\label{eq:stability-condition}
\end{equation}
On integrating by parts, the above equation gives the equality $\pi 
\int {\rm d}p~f_0(p)/(p-i\lambda)^2=i$ that can never be
satisfied for $\lambda$ purely imaginary. We thus conclude that
$\lambda$ is complex in general. 

Let us first consider the so-called waterbag (WB)
distribution, commonly used in studying long--range Hamiltonian
systems~\cite{Campa2014} and inspired by plasma physics, where the
particle momenta are uniformly distributed in a range $[-\Delta p: \Delta
p]$, with $\Delta p\geq 0$. The stability equation
\eqref{eq:stability-condition} translates into $\lambda^2 = \pm i/2-\Delta p^2$,
which shows that if $\lambda$ solves the above equation, so does
$-\lambda$, yet $\lambda$ cannot be pure imaginary. Thus, the
stability equation will always admit a solution with positive real part,
so that the WB distribution cannot be linearly stable under the dynamics~\eqref{eq:Vlasov}. 

We now consider a Gaussian state uniform in $\theta$ and Gaussian in
$p$: 
$f_0(p)=1/(2\pi\sqrt{2\pi\sigma^2})\exp(-p^2/(2\sigma^2))$, with
$\sigma>0$. Equation~\eqref{eq:stability-condition} gives 
$$\pm \frac{i}{\sqrt{\pi}(2\sigma^2)^{3/2}}\Big(\sqrt{2\pi\sigma^2}-\pi
\lambda
e^{\lambda^2/(2\sigma^2)}\mathrm{Erfc}\Big(\lambda/\sqrt{2\sigma^2}\Big)\Big)=1,$$
where $\mathrm{Erfc}(x)$ is the complementary error function. We have checked numerically that the above equation does not admit 
eigenvalues $\lambda$ with a non--negative real part, for any value of
$\sigma$, and hence we can conclude that a state Gaussian in $p$ and
uniform in $\theta$ is always stable under the Vlasov
dynamics~\eqref{eq:Vlasov}. Let us however remember that the
condition~\eqref{eq:stability-condition} is quite general, so some
non--Gaussian distributions may also be stable. For example, for a
Lorentzian distribution $f_0(p)=\sigma/\pi(p^2+\sigma^2)$, the eigenvalues are $\lambda=-\sigma\pm
\sqrt{\pi/2}(1+i)$, so that the distribution is stable provided its width obeys $\sigma>\sqrt{\pi/2}$.

On the basis of the above discussion, and as confirmed numerically, the dynamics of a large system initially in a WB configuration relaxes to a Vlasov-stable stationary state on an
$N$-independent timescale. Yet, since the system does not possess a proper equilibrium, its convergence
to a Gaussian state (Boltzmann distribution if the system were Hamiltonian) is not granted.

\begin{figure}[!ht]
\centering
\includegraphics[width=8.0cm]{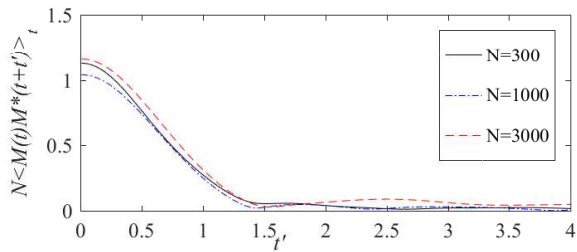}
\caption{(Color online) Auto-correlation in time of the magnetization $M(t)$, computed after a time $100N$, and over a time window $\Delta t=100$, starting from a WB initial state with $\Delta p=0.5$.}
\label{fig:m-autocorrelation}
\end{figure}

To understand the evolution of the Vlasov-stable distribution for finite
$N$, let us consider single particles: They are driven by the
magnetization, which fluctuates around zero. Using the definition of the
magnetization, let us rewrite the single--particle dynamics by using
Eq.~\eqref{eq:pjdot} as
\begin{equation}
\dot{p}_j=-\frac{1}{N}-\frac{\Re\big(\eta_j(t)\big)}{\sqrt{N}},
\label{eq:pjdotnoise}
\end{equation}
where $\eta_j(t)=e^{i\theta_j(t)}(1/\sqrt{N})\sum_{m\neq
j}e^{-i\theta_m(t)}$ is of order unity, and the factor $1/N$ comes from the
diagonal $j=m$ term in $M$. On timescales much
smaller than $\sqrt{N}$, the resulting quasi--ballistic motion makes it possible to write that $\theta_j(t+t')-\theta_j(t) \approx p_j(t)t'$. Assuming that the particles have uncorrelated positions, we obtain that $\dot{P}=-|M|^2=-1/N$ and 
\begin{eqnarray}
&&\langle\eta_j(t)\eta_j^\ast(t+t')\rangle \nonumber \\
&&\approx e^{-ip_j(t)t'}\frac{1}{N} \sum_{m\ne j} e^{ip_m(t)t'}
\left(1+ \sum_{n\neq m}e^{i(\theta_n(t)-\theta_m(t))}\right)\nonumber
\\ && \approx e^{-ip_j(t)t'}\iint {\rm d}\theta {\rm d}p~f_t(p) e^{ipt'},
\end{eqnarray}
where the double sum has been dropped in going from the second to the third line.
Here,
$\langle.\rangle$ represents an average over configurations, and $f_t$
the statistical average of the single-particle distribution $f$ at time
$t$. For a Gaussian distribution $f_t=1/(2\pi)\exp(-(p-\bar{p})^2/(2\sigma^2))/\sqrt{2\pi \sigma^2}$ centered around $\bar{p}$, one obtains
\begin{equation}
\langle\eta_j(t)\eta_j^\ast(t+t')\rangle=\exp{\left(-\frac{\sigma^2t^2}{2}-i(p_j-\bar{p})t'\right)}.\label{eq:eta2t}
\end{equation}
The phase term in Eq.~\eqref{eq:eta2t} may be neglected since it varies
little over the different values of $p_j$ (i.e., over the momentum
distribution) for times smaller than the coherence time $t'<1/\sigma$.
Consequently, for timescales larger than $1/\sigma$, $\eta_j$ can
effectively be considered as a white noise with $\langle
\eta(t)\rangle=0, ~\langle\eta(t)\eta(t+t')\rangle=D(\sigma)\delta(t')$,
where the diffusion coefficient is obtained as 
\begin{equation}
D(\sigma)=\int \langle\eta_j(t)\eta_j(t+t')\rangle dt=\sqrt{\frac{\pi}{2}}\frac{1}{\sigma}.\label{eq:Dgauss}
\end{equation}
This behavior of the magnetization is illustrated in
Fig.~\ref{fig:m-autocorrelation}, where the auto-correlation in time of
the magnetization is shown, presenting a clear decay in time over a
scale that does not depend on the system size $N$.
Moreover, this allows to write a Fokker-Planck equation for the single--particle
distribution ${\cal P}(p-P,t)$ centered around $P$ as
\begin{equation}
\frac{\partial {\cal P}}{\partial t}=D(\sigma)\frac{\partial^2 {\cal P}}{\partial p^2}.
\end{equation}
While this appears to be the equation of a Brownian motion, the
dependence of the diffusion coefficient on the distribution makes it a
{\it nonlinear} equation in ${\cal P}$, which does not possess an
analytical solution~\cite{nonlinear-FP}. Practically, as the
distribution spreads in momentum, the diffusion coefficient decreases as
the coherence time of $\eta(t)$ reduces, so that the diffusion actually
slows down in time.

Before describing the above process in more detail, let us comment on
the complete dynamical evolution starting from the initial WB state:
After the initial transient that follows the relaxation from the WB
state on a timescale that does not depend on the system size (a process
often called violent relaxation~\cite{LyndenBell1967}), the system reaches a
state that statistically corresponds to a distribution which is a
stationary and stable solution of the Vlasov equation. After that, the
slow (quasi-stationary) relaxation occurs over timescales that grow
linearly with the system size $N$, during which the system evolves
toward a state Gaussian in momentum and homogeneous in $\theta$. This was checked numerically by
monitoring the momenta of the distribution, which reached the
values for a Gaussian distribution. The dynamical evolution is shown in Fig.~\ref{fig:p-profiles}.

\begin{figure}[!ht]
\centering
\includegraphics[width=8cm]{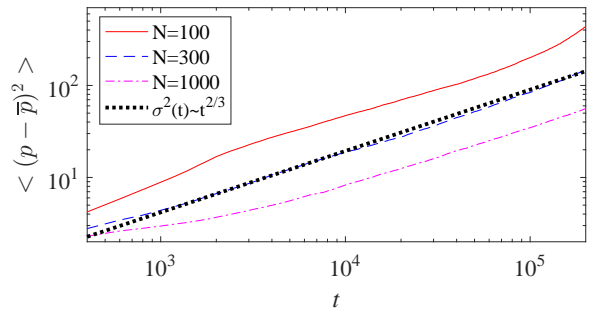}
\caption{(Color online) Evolution of $\langle(p-\bar{p})^2\rangle$ for
different system sizes, and the subdiffusive behavior prediction, Eq.~\eqref{eq:sigma}. The system is initially in a WB state with $\Delta p=0.5$.}
\label{fig:subdiffusion}
\end{figure}

The evolution of the distribution is then captured under the hypothesis
that it is Gaussian at any time. Using the ansatz ${\cal
P}(t)=1/(2\pi)\exp(-(p-\bar{p})^2/(2\sigma^2(t)))/\sqrt{2\pi \sigma^2(t)}$
along with Eq.~\eqref{eq:Dgauss}, one obtains $\sigma'\sigma=D$, which
yields
\begin{equation}
\sigma^3(t)=\sigma^3(0)+3\sqrt{\frac{\pi}{2}}t.\label{eq:sigma}
\end{equation}
This equation describes a {\it subdiffusive} behavior, where the
distribution temperature $T\sim \langle(p-\bar{p})^2\rangle$ grows with
time as
$t^{2/3}$, instead of $t$ as for the standard Brownian motion, due to the
fact that the spreading of the distribution in momentum continues
concomitantly with a reduction of the diffusion coefficient
\eqref{eq:Dgauss}. The validity of the Gaussian distribution ansatz is
confirmed by the numerical observation of the subdiffusive behavior, see
Fig.~\ref{fig:subdiffusion}.
This result bears strong similarities with those of
Ref.~\cite{Bouchet2005}, where anomalous diffusion was predicted for a
similar infinite--range {\it Hamiltonian} system. In that case, the
diffusion in the system was also resulting from the weak coupling of many particles through a vanishing magnetization.

In conclusion, we have shown that a non-Hamiltonian long-range system
may present a slowdown of relaxation with the system size, similar to
what is known for Hamiltonian systems under the name quasi--stationary
states. The existence of a Vlasov equation for non--conservative systems
driven by non--Hamiltonian two-body
interaction (differently from, for
example, systems with friction forces) allows for this
approach to possess non--equilibrium stable stationary states, which
translates into quasi--stationary states for the microscopic dynamics. The increase over time of the system temperature turns the interaction
between the particles less and less effective. Because each particle in
the non--magnetized phase feels the coupling to all other particles
through an effective noise, this results in a diminishing diffusion constant and a subdiffusive behavior.

A particularly promising platform to investigate experimentally the
aforementioned peculiar behavior is that of an ultracold cloud trapped in an optical
cavity. In this case, the infinite--range interaction between the atoms
mediated by the light is known to dominate the dynamics, and the leakage
of the light through the cavity mirrors results in an overdamped
dynamics. These systems do not have a thermal equilibrium state, since the pump
light keeps increasing the cloud momentum, driving the atoms farther and
farther from resonance. The fact that the momentum distribution is
routinely tracked by time-of-flight techniques make these setups
especially interesting for observing the predicted non--equilibrium anomalous diffusive behavior.

\textit{Acknowledgements:} This paper was written up during SG's visit to the Universidade Federal de S\~{a}o Carlos and the Centro de Pesquisa em
\'{O}ptica e Fot\^{o}nica (FAPESP 2013/07276-1), Brazil during June 2018. He thanks these institutions for warm hospitality and financial support. RB
hold grants from Funda\c{c}\~ao de Amparo \`a Pesquisa do Estado de S\~ao Paulo (FAPESP) (2014/01491-0 and 2015/50422-4). NP and RB participate
in the EU H2020 ITN project ColOpt (No. 721465).  The Titan X Pascal
used for this research was donated by the NVIDIA Corporation.
\bibliography{../../Biblio/BiblioCollectiveScattering}

\end{document}